\documentclass{jfm}

\usepackage{graphicx}
\usepackage{amssymb}
\usepackage{amsfonts}
\usepackage{natbib}
\usepackage{pstool}
\usepackage{xcolor}
\usepackage{verbatim}
\usepackage{pifont}
\usepackage{psfrag}

\title[Reynolds stress scaling in the near-wall region]{Reynolds stress scaling in the near-wall region}

\author[Smits and Hultmark]%
{A\ls
L\ls E\ls X\ls A\ls N\ls D\ls E\ls R\ns J.\ns S\ls M\ls I\ls T\ls S \and  M\ls A\ls R\ls C\ls U\ls S\ns H\ls U\ls L\ls T\ls M\ls A\ls R\ls K }

\affiliation{Department of Mechanical and Aerospace Engineering, 
Princeton University, \break
Princeton, NJ 08544, USA\\[\affilskip]
}

\pubyear{2021}
\volume{}
\pagerange{}

\begin{document}

\maketitle

\begin{abstract}
A new scaling is derived that yields a Reynolds number independent profile for all components of the Reynolds stress in the near-wall region of wall bounded flows.  The scaling demonstrates the important role played by the wall  shear stress fluctuations and how the large eddies determine the Reynolds number dependence of the near-wall turbulence behavior. 
\end{abstract}

\section{Introduction}
\label{intro}

Here, we examine the near-wall scaling behavior of turbulent flows on smooth walls.   These flows include two-dimensional zero-pressure gradient boundary layers, and fully developed pipe and channel flows.  The focus is on the region $y^+<100$, which includes the peaks in the streamwise and spanwise turbulent stresses.  The superscript $^+$ denotes non-dimensionalization using the fluid kinematic viscosity  $\nu$ and the friction velocity $u_\tau=\sqrt{\tau_w/\rho}$, where $\tau_w$ is the mean wall shear stress and $\rho$ is the fluid density.   

\begin{figure}
\centering
\includegraphics[width=0.47\textwidth]{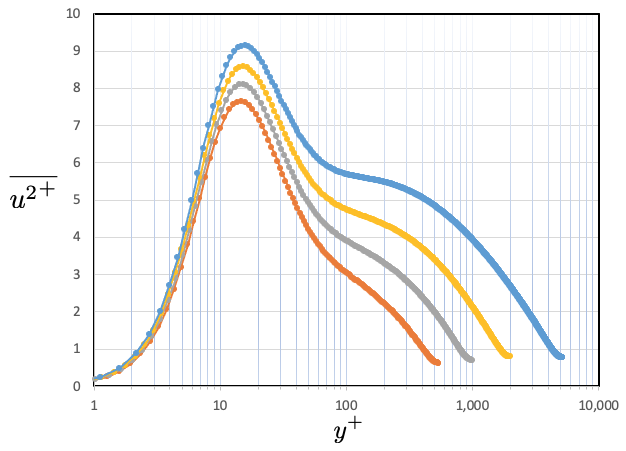} 
\includegraphics[width=0.47\textwidth]{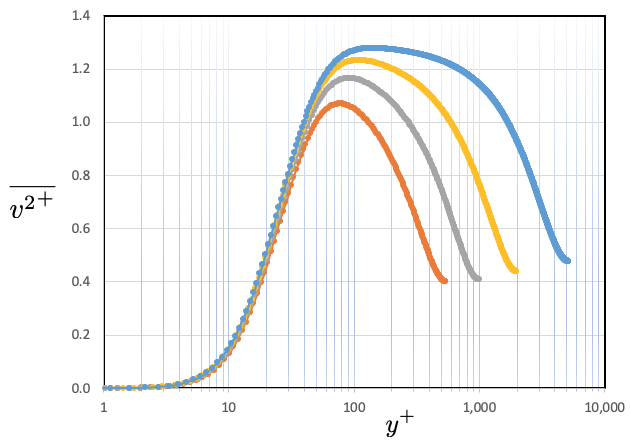} \\
\includegraphics[width=0.47\textwidth]{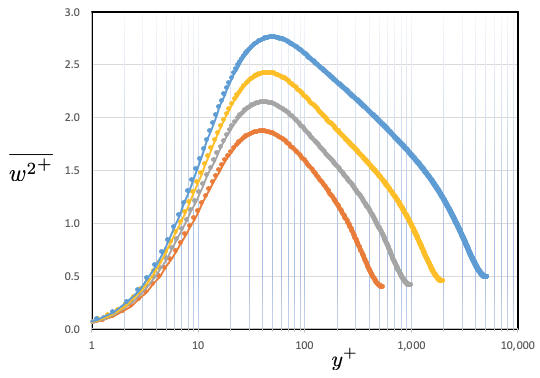} 
\includegraphics[width=0.47\textwidth]{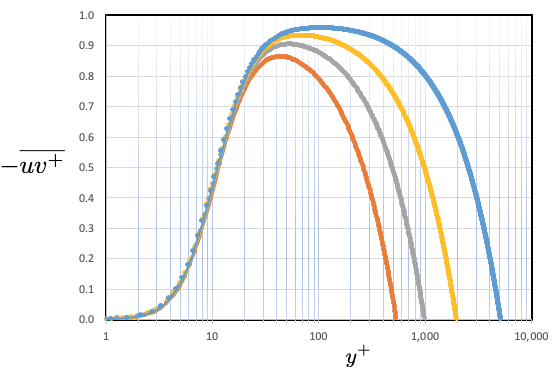} 
\caption{ Profiles of turbulent stresses in channel flow, as computed by DNS  $Re_\tau =$ 550, 1000, 2000, 5200 \citep{lee2015}.  }
\label{all_stresses}
\end{figure}

For isothermal, incompressible flow, it is commonly assumed that for the region close to the wall
$$[U_i,\overline{({u_i u_j})^+}] = f(y, u_\tau, \nu, \delta),$$
where $U_i$ and $u_i$ are the mean and fluctuating velocities in the $i$th direction.  The overbar denotes ensemble averaging, and the outer length scale $\delta$ is, as appropriate, the boundary layer thickness, the pipe radius, or the channel half-height.  That is,
\begin{equation}
[U_i^+,\overline{({u_i u_j})^+}] = f(y^+, Re_\tau).
\label{dim_analysis}
\end{equation}
where the friction Reynolds number $Re_\tau=\delta u_\tau/\nu$.   The Reynolds number dependence of $\overline{u^2}$ (streamwise direction), $\overline{v^2}$ (wall-normal direction), $\overline{w^2}$ (spanwise direction), and $-\overline{uv}$ (Reynolds shear stress) is illustrated in figure~\ref{all_stresses}.

\cite{hultmark2021smits} considered the scaling of the streamwise component $\overline{u^2}$.  Specifically, they showed that a new velocity scale, defined by $b_1 u_\tau$ collapsed the profiles of $\overline{u^2}$ onto a single curve for the whole of the region $y^+ < 20 $ (including the inner peak in $\overline{u^2}$).  That is,
\begin{equation}
\overline{{u^2}^+} = \overline{b_1^2} (Re_\tau) f(y^+).
\label{fg}
\end{equation}
The parameter $\overline{b_1^2}$ scales with the mean square value  of the fluctuating wall shear stress $\tau_w'$, and it was derived using a Taylor Series expansion of the fluctuating velocity very close to the wall.  The analysis used the channel flow direct numerical simulations (DNS) by \cite{lee2015}, with strong support from high Reynolds number experimental data by \cite{Vallikivi2015_HRTF} and \cite{samie2018fully}.  Here, we generalize this scaling to include the other components of the Reynolds stress tensor $\overline{v^2}$ (wall-normal direction), $\overline{w^2}$ (spanwise direction), and $-\overline{uv}$ (Reynolds shear stress).

\section{Taylor Series expansions}

We start in the same way as \cite{hultmark2021smits}  by writing the Taylor series expansions for $u_i$ in the vicinity of the wall.  Instantaneously \citep{bewley2004skin},
\begin{eqnarray}
\underline u^+ & = & a_1 + b_1 y^+  + c_1{y^+}^2 + O({y^+}^3) \label{Taylor1u} \\
\underline v^+ & = & a_2 + b_2 y^+ + c_2{y^+}^2 + O({y^+}^3) \label{Taylor1v} \\
\underline w^+ & = & a_3 +  b_3 y^+ +  c_3{y^+}^2 + O({y^+}^3) \label{Taylor1w}
\end{eqnarray}
where $\underline u=U+u$, etc.  The no-slip condition gives $a_1=a_2=a_3=0$, and by continuity $\partial v /\partial y|_w = b_2 = 0$.
Also 
\begin{eqnarray}
b_1 & = &  (\partial \underline  u^+/ \partial y^+)_w,  \label{b1}  \\
b_3 & = &  (\partial \underline w^+/ \partial y^+)_w,  \label{b3} \\
c_2 & = & -{\textstyle {1 \over 2}}  (\partial b_1/ \partial x^+ + \partial b_3/ \partial z^+).   \label{c2}
\end{eqnarray}

For the time-averaged quantities
\begin{eqnarray}
\overline{{u^2}^+} & = & \overline{{b_1}^2}{y^+}^2 + \dots \label{Taylor1u2} \\
\overline{{v^2}^+} & = &  \overline{{c_2}^2}{y^+}^4 + \dots \label{Taylor1v2} \\
\overline{{w^2}^+} & = & \overline{{b_3}^2}{y^+}^2 + \dots \label{Taylor1w2} \\
-\overline{{uv}^+} & = & \overline{b_1 c_2}{y^+}^3 + \dots  \label{Taylor1uv}
\end{eqnarray}

\begin{figure}
\centering
\includegraphics[width=0.45\textwidth]{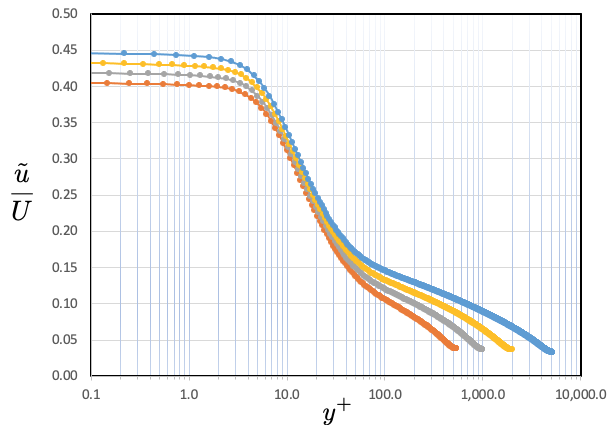} \hspace{3mm}
\includegraphics[width=0.47\textwidth]{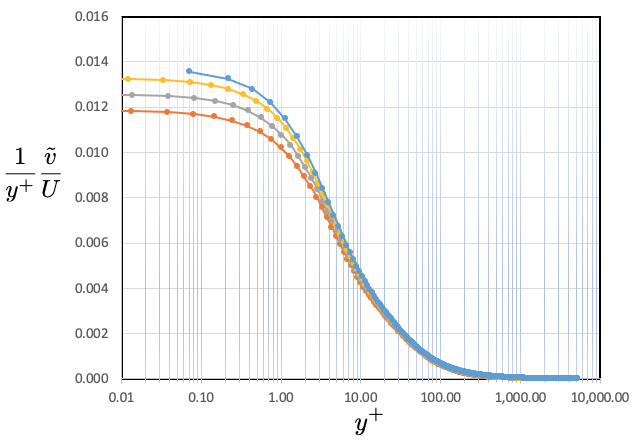} \\
\includegraphics[width=0.45\textwidth]{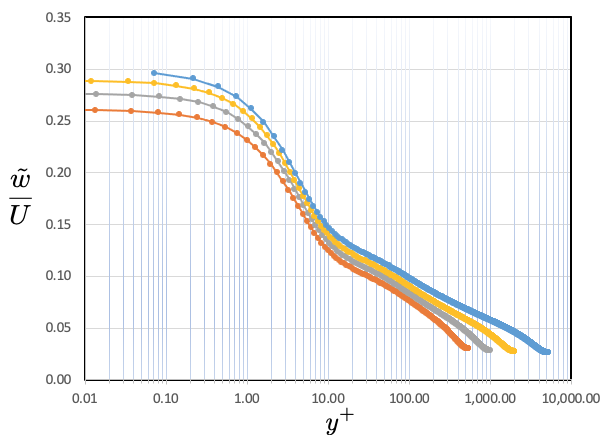}
\includegraphics[width=0.51\textwidth]{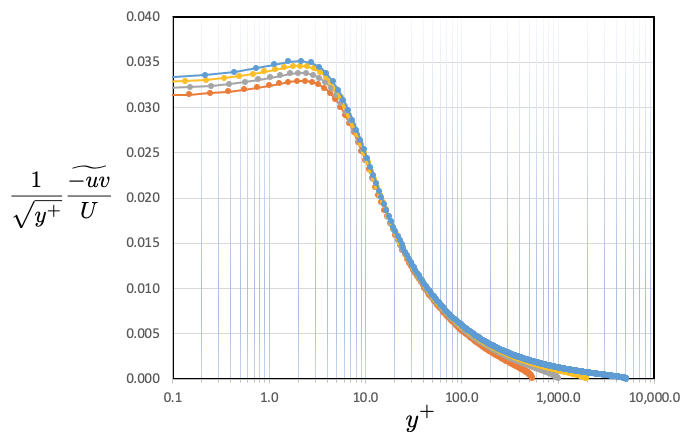}
\caption{ Top left: Profiles of $\tilde u/U$; intercept at $y^+=0$ is $\tilde b_{1}$.  Top right: Profiles of $(\tilde v/U)/y^+$; intercept at $y^+=0$ is $\tilde c_{2}$.  Bottom left: Profiles of $\tilde w/U$; intercept at $y^+=0$ is $\tilde b_{3}$.  Bottom right: Profiles of $(\widetilde{-uv}/U)/\sqrt{y^+}$; intercept at $y^+=0$ is $\widetilde {b_{1} c_2}$.  $Re_\tau =$ 550, 1000, 2000, 5200 \citep{lee2015}.  }
\label{u_U_DNS}
\end{figure}

These expansions can only hold very close to the wall.  Since the mean velocity is linear in this region 
\begin{eqnarray}
\tilde u/U & = & \tilde b_{1} \label{Taylor_rms_u} \\
\tilde v/U & = & \tilde c_{2} y^+ \label{Taylor_rms_v} \\
\tilde w/U & = & \tilde b_{3}  \label{Taylor_rms_w} \\
\widetilde{-uv}/U & = & \widetilde {b_{1} c_2} \sqrt{y^+}  \label{Taylor_rms_uv}
\end{eqnarray}
where $ \tilde u = \sqrt{\overline{{u^2}^+}}$, $ \tilde b_1 = \sqrt{\overline{b_1^2}}$, $ \widetilde {b_{1} c_2} = \sqrt{\overline{b_1c_1}}$, and $ \widetilde {u v} = \sqrt{\overline{u v}}$. 
The channel flow DNS data given in figure~\ref{u_U_DNS} shows that at a given Reynolds number $ \tilde b_1$, $ \tilde c_2$ , $ \tilde b_3$ and $ \widetilde {b_{1} c_2}$ all become constant as the wall is approached.

\begin{figure}
\centering
\includegraphics[width=0.5\textwidth]{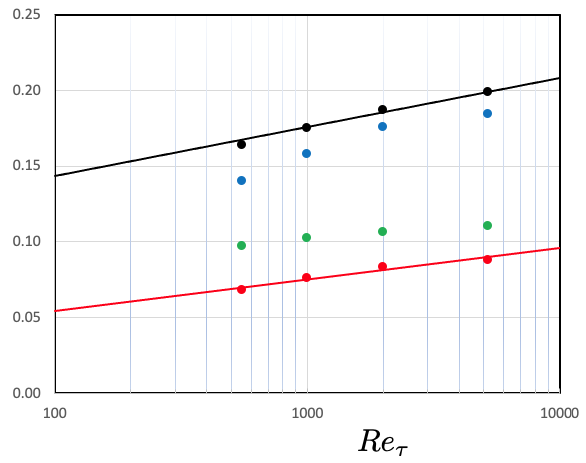} \caption{ Reynolds number variation of $\overline{{b_1}^2}$ (black); $1000\overline{{c_2}^2}$ (blue); $\overline{{b_3}^2}$ (red);  $-100\overline{b_1c_2}$ (green).  See equations~\ref{constants2_log_b1} to \ref{constants2_log_b3},  Channel flow DNS \citep{lee2015}.  }
\label{constants}
\end{figure}

The Reynolds number dependencies of $\overline{{b_1}^2}$,  $\overline{{c_2}^2}$, $\overline{{b_3}^2}$, and $\overline{b_1 c_2}$ are shown in figure~\ref{constants}, where it could be argued that $\overline{{b_1}^2}$ and $\overline{{b_3}^2}$ for this range of Reynolds numbers the variation with Reynolds number is close to logarithmic, with 
\begin{eqnarray}
\overline{{b_1}^2} & = & 0.0796 + 0.0139 \ln{Re_\tau} \label{constants2_log_b1} \\
\overline{{b_3}^2}  & = & 0.012 + 0.00091 \ln{Re_\tau}  \label{constants2_log_b3}
\end{eqnarray}
Note that $-\overline{b_1 c_2}$ cannot increase without limit since  $-\overline{uv}/u_\tau^2$ cannot exceed one in channel flow; $\overline{c_2^2}$ must therefore also be limited.  In addition, \cite{chen2021reynolds}, by considering the boundedness of the near-wall dissipation rate, argued that the logarithmic increase in $\overline{b_1^2}$ cannot hold for all Reynolds numbers. 

From equations~\ref{b1} and \ref{b3}
\begin{equation}
\overline{b_1^2} =  \frac{\overline{{\tau'_{wx}}^2}}{\tau_w^2}  \quad \hbox{and} \quad  \overline{b_3^2} = \frac{\overline{{\tau'_{wz}}^2}}{\tau_w^2}. 
\label{uadef}
\end{equation}
The controlling parameter in the near-wall scaling for $u$ is therefore the mean square of the fluctuating wall stress in the $x$-direction $\tau'_{wx}$, and for $w$ it is the mean square of the fluctuating wall stress in the $z$-direction $\tau'_{wz}$.   It is more difficult to give a precise meaning to $c_2$ and $b_1c_2$, but they express correlations between spatial gradients of the instantaneous wall stress fluctuations, as well as the fluctuating wall stress itself. It is therefore to be expected that $\tilde c_2$ and $\widetilde{b_1c_2}$ are more connected to the small-scale motions than either $\tilde b_1$ or $\tilde b_3$.

As pointed out by \cite{hultmark2021smits}, this connection with the fluctuating wall stress helps explain the Reynolds number dependence seen in figure~\ref{constants} since with increasing Reynolds number the large-scale (outer layer) motions contribute more and more to the fluctuating wall stress.  Because $\tilde c_2$ and $\widetilde{b_1c_2}$ are more connected to the small-scale motions than either $\tilde b_1$ or $\tilde b_3$, it might be expected that they feel the effects of modulation more than superimposition by the large-scale motions \citep{Marusic_Science}.

\section{Scaling the $\widetilde{{u_1u_j}^+}$ profiles}

\begin{figure}
\centering
\includegraphics[width=0.46\textwidth]{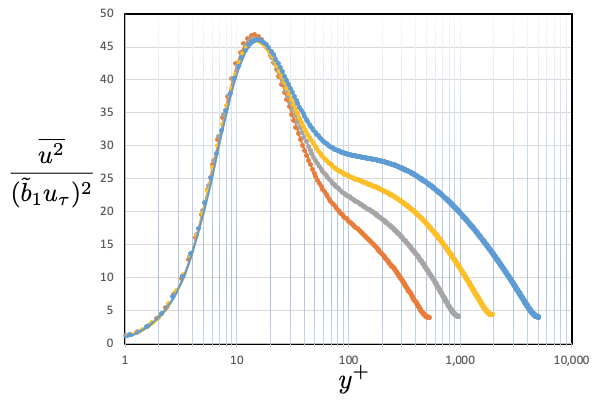}  \hspace{1mm}
\includegraphics[width=0.51\textwidth]{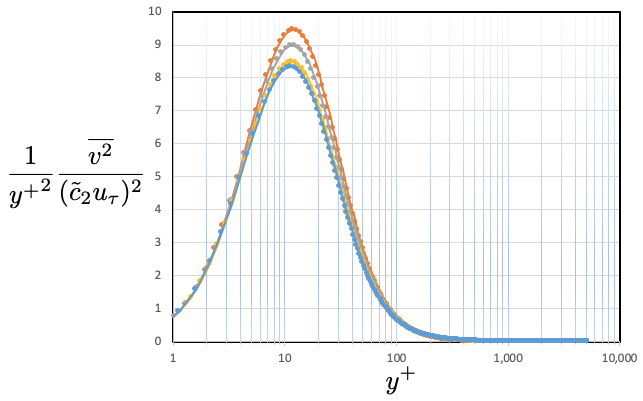} \\
\includegraphics[width=0.46\textwidth]{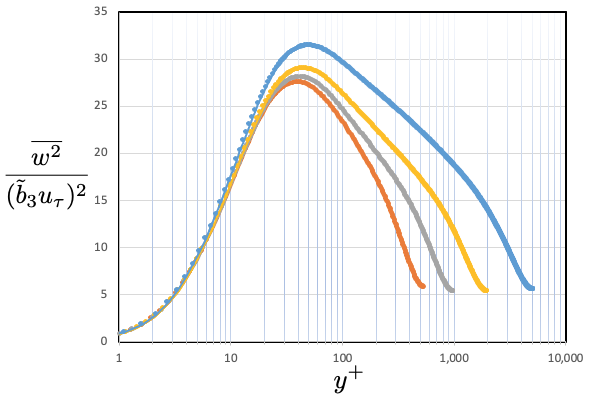} \hspace{1mm}
\includegraphics[width=0.51\textwidth]{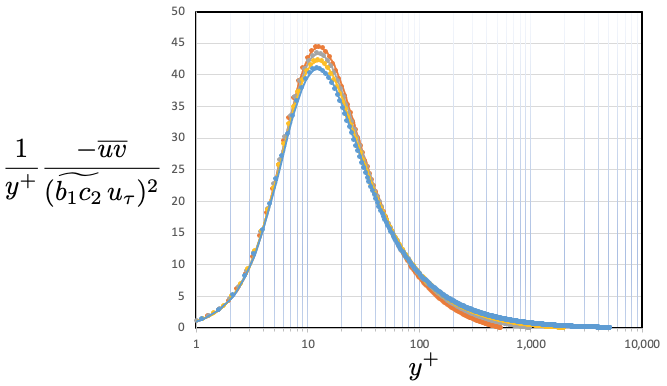}
\caption{ Profiles of scaled stresses for $Re_\tau =$ 550, 1000, 2000, 5200.   Channel flow DNS \citep{lee2015}.  }
\label{scaled_stresses}
\end{figure}

\cite{hultmark2021smits} showed that when each $\overline{{u^+}^2} $ profile is scaled with the value of $\overline{b_1^2}$ at its respective Reynolds number, we obtain the results shown in figure~\ref{scaled_stresses}.  The collapse of the data for $y^+<20$ is remarkable, including the almost perfect agreement on the scaled peak value $\overline{{u^2_p}^+}$.  Furthermore, $\overline{{b_1}^2}$ and $\overline{{u_p^2}^+}$ approach a constant ratio to each other such that for channel flow,
\begin{equation}
\overline{{u_p^2}^+} \approx 46\overline{{b_1}^2}.
\label{u2b12}
\end{equation} 
  In other words, the magnitude of the peak at $y^+ \approx 15$ tracks precisely with $\overline{{b_1}^2}$, a quantity that is evaluated at $y^+=0$.  A similar conclusion was also proposed by \cite{chen2021reynolds}.

The results for scaling the other components of the stress tensor in a similar way are also shown in figure~\ref{scaled_stresses}. The collapse of the data is not quite as impressive as for the streamwise component, but it is nevertheless encouraging, especially when compared to the original unscaled data shown in figure~\ref{all_stresses}.

\begin{figure}
\centering
\includegraphics[width=0.45\textwidth]{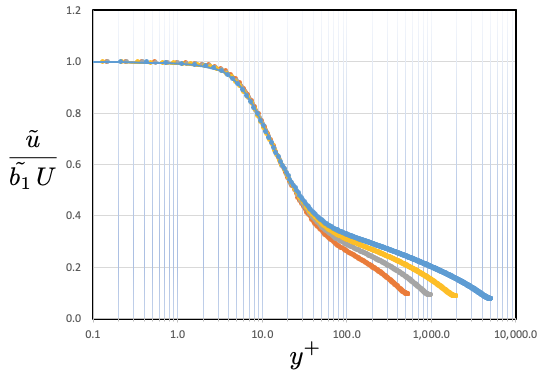} \hspace{2mm}
\includegraphics[width=0.49\textwidth]{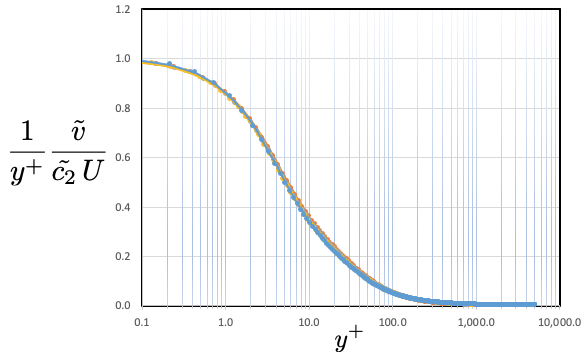} \\
\includegraphics[width=0.46\textwidth]{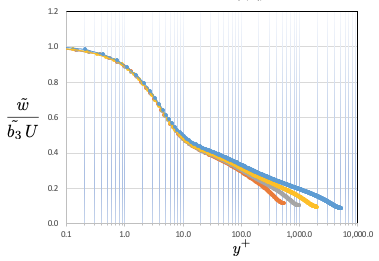}
\includegraphics[width=0.51\textwidth]{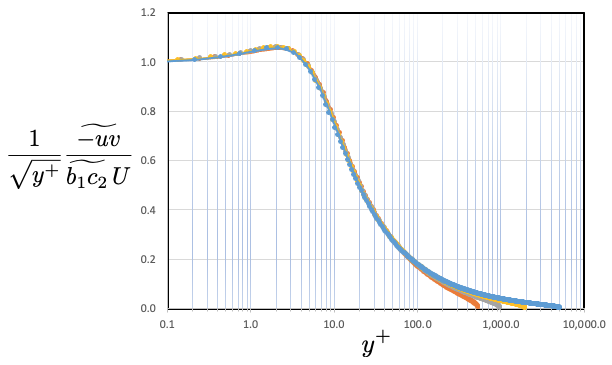}
\caption{ Profiles of scaled local stresses.  $Re_\tau =$ 550, 1000, 2000, 5200 \citep{lee2015}.  }
\label{u_v_U_scaled_DNS}
\end{figure}

If we now return to using the normalization by the local mean velocity, originally given in figure~\ref{u_U_DNS}, and apply the scaling suggested by equations~\ref{Taylor_rms_u} to \ref{Taylor_rms_uv} we obtain the results shown in figure~\ref{u_v_U_scaled_DNS}.  For $\tilde u$ and $\tilde w$, the collapse holds for $y^+<20$ and $y^+<10$, respectively, which is comparable to that seen for the scaling without $U$, as given in figure~\ref{scaled_stresses}.  However, for $\tilde v$, and $\widetilde{-uv}$ the general collapse is much improved, and it extends as far out from the wall as $y^+=  100$.  

\section{Conclusions}

By expanding the velocity in a Taylor Series with distance from the wall, the Reynolds number dependence of the near-wall distributions of the Reynolds stresses was traced to the magnitude of the fluctuating wall shear stress and its spatial gradients, which are increasingly affected by the superimposition and modulation of the near-wall motions due to large-scale, outer-layer motions as the Reynolds number increases \citep{Marusic_Science}.  

The Taylor series expansion also suggests a separate scaling for each component of the Reynolds stress.  For the streamwise component, its scaling collapses the data for $y^+ < 20$, a region that includes the near-wall peak, as shown earlier by \cite{hultmark2021smits}.  For the spanwise component, the scaling holds only up to about $y^+ < 10$, and fails to capture the growth of its inner peak.  For the wall-normal component and the Reynolds shear stress, the proposed scaling uses the local mean velocity rather than the friction velocity (as suggested by the analysis), and gives a very satisfactory collapse up to about $y^+=100$.  Hence, for the stresses that depend on the wall-normal component of the velocity fluctuation, the near-wall behavior is governed both by the fluctuating wall shear stress and its spatial gradients, as well as the local mean velocity.  

Revisiting the dimensional analysis given in equation~\ref{dim_analysis}, we can now be more precise and write for the Reynolds stresses in a two-dimensional wall-bounded flow
\begin{equation}
{\widetilde{u_i u_j}}/U = g_1(Re_\tau) g_2(y^+),
\label{dim_analysis2}    
\end{equation}
where $g_1$ and $g_2$ and their ranges of applicability are different for each component.  Specifically, with $U^+=f(y^+)$ absorbed into $g_2$, a consistent framework for all component is given by
\begin{eqnarray*}
\tilde u: & \quad & g_1=\tilde b_1  \hspace{8mm}  (y^+<20); \\
\tilde v: & \quad & g_1=\tilde c_2  \hspace{8mm}  (y^+<100); \\
\tilde w: & \quad & g_1=\tilde b_3  \hspace{8mm}  (y^+<10); \\
-\widetilde{uv}: & \quad & g_1=\widetilde{b_1c_2}  \hspace{5mm}  (y^+<100).
\end{eqnarray*}

Our conclusions were supported only by channel flow DNS.  Although we anticipate that they would extend to boundary layer flows (as Hultmark \& Smits 2021 showed for the streamwise component) and fully developed pipe flows, high Reynolds number DNS for such flows will be needed to confirm these expectations.   

It may also be remarked that because the scaling is different for each component of the stress, any isotropic definition of eddy viscosity will obviously fail in the near-wall region.  The use of an eddy viscosity in this region is further invalidated by the important role of the mean velocity in scaling the stresses that depend on the wall-normal component of the velocity fluctuation, rather than its gradient.  In this respect, \cite{Hultmark2013a} noted that in the overlap region, where both the mean velocity $U$ and the streamwise stress $\overline{{u^2}^+}$ follow a logarithmic distribution, $\overline{{u^2}^+}$ depends on $U$ rather than its gradient.

\subsection*{Acknowledgments}
This work was supported by ONR under Grant N00014-17-1-2309 (Program Manager Peter Chang). We thank Myoungkyu Lee and Robert Moser for sharing their data (available at http://turbulence.ices.utexas.edu). 

\bibliographystyle{jfm}
\bibliography{BigBib_master_12_15_20}

\end{document}